\documentclass{blois}

\def\be{\begin{equation}}
\def\ee{\end{equation}}
\def\bea{\begin{eqnarray}}
\def\eea{\end{eqnarray}}

\newcommand{\Photo}{}

\begin{document}
\vspace*{4cm}
\title{$B$-PHYSICS ANOMALIES: UV MODELS}

\author{J. M. LIZANA}

\address{Physik-Institut, Universit\"at Z\"urich, Winterthurerstrasse 190, \\
CH-8057 Z\"urich, Switzerland}

\maketitle\abstracts{$B$-anomalies hint towards New Physics violating Lepton Flavor Universality at the TeV scale. The way they manifest points out to very particular structures for this New Physics. In this talk I review some UV-complete models to explain the $B$-anomalies, and possible directions these models suggest for Physics beyond the TeV scale.
}

\section{Introduction}

In the recent years, hints towards New Physics (NP) manifesting Lepton Flavor Universality Violation (LFUV) have been growing. The semileptonic decays of $B$ mesons deviate from the Standard Model (SM) prediction by more than $4\,\sigma$ in the most conservative fits.\cite{Cornella:2021sby} These are the so-called $B$-anomalies. The observables that present deviations can be divided into two groups:
\begin{itemize}
\item {\it Neutral-current anomalies}: decays of $B$ mesons that at the partonic level correspond to $b\to sll$. They include the Lepton Flavor Universality (LFU) ratios $R_{X_s}$,
\begin{equation}
R_{X_s}=\frac{{\rm Br}(B\to X_s\mu\bar \mu)}{{\rm Br}(B\to X_s e\bar e)},
\end{equation}
where $X_s$ is any strange meson as $K$ or $K^*$,
branching ratios as $B_s\to \mu\bar \mu$ or $B\to X_sl\bar l$, and angular distributions in the decay $B\to K^*\mu\bar \mu$.
\item {\it Charge-current anomalies}: decays of $B$ mesons corresponding to $b\to c\tau \nu$ at the partonic level. The observables are the LFU ratios
\begin{equation}
R_{D^{(*)}}=\frac{{\rm Br}(B\to D^{(*)}\tau \bar\nu)}{{\rm Br}(B\to D^{(*)}\ell \bar\nu)},
\end{equation}
where $\ell$ denotes both $e$ and $\mu$.
\end{itemize}
Among them, $R_{K^{(*)}}$ constitutes one of the cleanest hints for NP. The cancelation of hadronic uncertainties in the ratio makes its theoretical calculation very robust. The current experimental value\,\cite{LHCb:2021trn} $R_K=0.864^{+0.044}_{-0.041}$  is $3.1\,\sigma$ away from the SM prediction, $1.00\pm 0.01$. This exact value is consequence of the particular structure of the SM: its gauge interactions have a $U(3)^5$ flavor symmetry, only broken by the Yukawa interactions between the Higgs boson and the fermionic fields. In addition, the Yukawa couplings have a highly hierarchical structure, so this breaking is specially suppressed for the first and second (light) families. 

There are strong experimental bounds on NP breaking the flavor symmetry, specially for the light families. In particular, $K-\bar K$ or $D-\bar D$ mixing, or bounds on $\mu \to e\gamma$ impose NP scale bounds of $\Lambda_{NP}>10^{4-5}\,$TeV. To overcome these constraints, one approach to study NP is the Minimal Flavor Violation (MFV), where {\it minimal} means that the only source of breaking of the flavor symmetry are the Yukawa couplings, at least up to some very high scale. Any explanation of the structure of this breaking would be postpone to this high scale. But $B$-anomalies change this picture. The only way to explain observables such as $R_K^{(*)}$ involves NP at the TeV scale with LFUV. In such scenario, we can still partially protect the strong flavor bounds reducing the minimally broken flavor symmetry to $U(2)^5$ for the light families. 
NP at the TeV scale could be strongly coupled to the third generation, opening the possibility of finding connections between apparently disconnected problems as the Higgs hierarchy problem, $B$-anomalies, and the flavor structure of the third family and its mixing with light families.

\section{Towards a combined UV explanation}

In the search of an explanation for $B$-anomalies through a UV complete model, we should first choose which mediator will give the anomalous contribution to the $B$ decays. 
The neutral-current anomalies, and in particular $R_{K^{(*)}}$, could be explained with a heavy $Z^{\prime}$ with a Flavor Changing Neutral Current (FCNC) between $b$ and $s$ and non-universal couplings to $\mu$ and $e$.\cite{Capdevila:2020rrl} However, $B_s$-mixing requires the $\bar b {Z}^{\prime} s$ coupling to be suppressed,
which needs to be compensated by a large coupling to muons $O(1)$, different to the coupling to electrons. This solution will strongly break the light-family flavor symmetry in the lepton sector, having only a minimally broken $U(2)^3$ symmetry in the quark sector. Some mechanism to suppress FCNC in the lepton sector must be additionally implemented in these models (for instance, extra symmetries).

A different explanation comes with the leptoquark (LQ): a color triplet charged under lepton number. It generates at tree level the semileptonic operators necessary to explain both neutral- and charge-current anomalies. At the same time, it only gives contribution to $\Delta F=2$ processes at one loop reducing considerably its $B_s$-mixing contribution, and its production at the LHC is highly suppressed avoiding direct searches constraints.
Different kinds of LQs have been studied in the literature.\cite{Angelescu:2021lln} In particular, the scalar LQ $S_3\sim (\bar {\bf 3},{\bf 3},1/3)$ or the vectors $U_1 \sim ( {\bf 3},{\bf 1},2/3)$ or $U_3\sim ( {\bf 3},{\bf 3},2/3)$ can explain $R_{K^{(*)}}$, and the scalars $S_1\sim (\bar {\bf 3},{\bf 1},1/3)$ or  $R_2\sim ({\bf 3},{\bf 2},7/6)$, or the vector $U_1$, can explain $R_{D^{(*)}}$.

\subsection{The $U_1$ leptoquark}

Among the different LQs mentioned above, only the vector $U_1\sim( {\bf 3},{\bf 1},2/3)$ can explain simultaneously both neutral- and charge-current anomalies. It also has the advantage of not generating $b\to s\nu\nu$ contributions at tree level, which would impose tight constraints on the model otherwise. The interaction terms we can write between $U_1$ and the SM fields read
\begin{equation}
\mathcal{L} \ni \frac{g_U}{\sqrt{2}}U_1^{\mu}\left[ \beta_L^{i\alpha} (\bar q_L^{\,i} \gamma_{\mu}l_L^{\alpha}+\beta_R^{i\alpha} (\bar d_R^{\,i} \gamma_{\mu}e_R^{\alpha})\right] +h.c.\label{eq:U1Lag}
\end{equation}
where I choose the basis where right-handed (RH) and down fields are in mass eigenstates. Fits of this model to the $B$-semileptonic-decay observables suggest the LQ should be strongly coupled to the third family and the left-handed (LH) current couplings should pay a $\sim 0.1$ suppression each time the family number of the fermions in the interaction is decreased, for instance, $\beta_L^{b\mu} \sim 0.1\, \beta_L^{b\tau}$. RH currents however can only be non-vanishing for $\beta_R^{b\tau}$ to suppress chirally enhanced contributions to $B_s\to\bar\mu\mu$. Therefore, this model can be protected with a minimal $U(2)^5$ breaking mechanism.

A $U_1$ LQ can appear as a composite resonance of a strongly coupled sector, as a Kaluza-Klein (KK) resonance in an extra dimension theory, or as a gauge boson getting a mass from a spontaneous symmetry breaking (SSB). I will focus in the last possibility first. Massive $U_1$ gauge bosons appear in quark-lepton unification theories, as the Pati-Salam (PS) model. In these models, $SU(3)_c$ is embedded into $SU(4)$, and fundamental fermions are identified with the SM fields as $\Psi=(q^{\,a},l)$ where $a=1,2,3$ is the color index. In particular, the PS gauge group is $PS=SU(4)\times SU(2)_L\times SU(2)_R$.
However, in the PS model all families are equally charged. Flavor bounds of the light families impose then strong limits on the mass of the LQ, $M_{LQ}>100\,$TeV, making the minimal version of this model nonviable for a $B$-anomaly explanation. 

\subsection{$4321$- models}

The constraints of the PS model can be overcome eliminating flavor universality. This is what is done in the so-called 4321 models, with the gauge group $SU(4)\times SU(3)^{\prime} \times SU(2)_L \times U(1)_X$ broken at the TeV scale to the SM gauge group by means of some 4321 SSB scalar fields. This breaking produces a massive $U_1$ together with a $Z^{\prime}$ and a color octet $G^{\prime}$ that can have non-universal couplings to the SM families depending on the charges of the fields. 
In addition, it will be necessary to add vector-like fermions (VLF) in a $({\bf 4},{\bf 1},{\bf 2},0)$ representation that can mix with the SM doublets charged under $SU(3)^{\prime}$ through Yukawa couplings with the 4321 SSB scalars. This mixing can be used to get chiral SM doublets partially charged under $SU(4)$ in the SM symmetric phase. In the following I define the $SU(3)$ ($SU(4)$)-site as the set of fields charged under $SU(3)$ ($SU(4)$). 
There are two possible charge assignments to get the structure of Eq.~\ref{eq:U1Lag}:
\begin{itemize}

\item {\it Universal 4321}:\,\cite{DiLuzio:2017vat} All SM fields are initially charged under $SU(3)^{\prime}$ and then, the third-family doublets are rotated with the VLFs to locate them mostly on the $SU(4)$-site. The second-family doublets also need to be slightly rotated to the $SU(4)$-site to couple them to the LQ and induce mixings with the third family. This setup can correctly reproduce the necessary structure for the LH currents of $U_1$ of  Eq.~\ref{eq:U1Lag}, and it makes all RH currents vanish because no RH SM fields are charged under $SU(4)$. In this model, the CKM mixing between heavy and light families is generated via coupling terms in the $SU(3)$-site, as opposite to the LQ mixing, which is realized in the $SU(4)$-site.

\item {\it Non-universal 4321}:\,\cite{Bordone:2017bld,Greljo:2018tuh} Light SM families are charged under $SU(3)^{\prime}$ and the third family under $SU(4)$. Second-family doublets are again rotated to the $SU(4)$-site with a mixing angle $O(0.1)$. With the appropriate mixing with the third-family doublets, LH currents can reproduce Eq.~\ref{eq:U1Lag}. There are RH currents only with the third family, as RH fields of the third family are charged under $SU(4)$ to cancel gauge anomalies. In addition, RH rotations between third and light families are suppressed, so the structure of $\beta_R^{i\alpha}$ in Eq.~\ref{eq:U1Lag} is preserved.
In this model, CKM and LQ mixings between third and light families are generated via coupling terms in the $SU(4)$-site, so both mixings are related.
\end{itemize}

\section{Beyond 4321}

Both models considered above have a rich phenomenology at the TeV scale testable via multiple searches in the near future. Although they are already UV-complete models, they bring new questions that invite to look for models beyond 4321. The fact that the NP appears at the TeV scale suggests there could be some connection with the Higgs hierarchy problem, as its proposed solutions typically also involve NP at the TeV scale. In this line, it has been studied models where the SSB of 4321 gauge symmetry is triggered by a condensate of a strongly coupled sector with a global symmetry $SU(4)_l\times SU(4)_h\times G_{EW}$, where the Higgs boson appears as a pseudo Nambu-Goldstone boson emerging from this sector as in composite Higgs models.\cite{Fuentes-Martin:2020bnh}

The quark-lepton unification of 4321 models suggests that light families could also unify at higher scales in a non-universal way. This is achieved by the $PS^3$ model,\cite{Bordone:2017bld} a three-site model with a different PS group and one family on each site. This gauge group is broken through link fields between the sites at different scales in cascade. The breaking of PS would be more localized on the light-family sites, while the Higgs and the breaking of the electroweak (EW) symmetry on the third-family site. This would explain the flavor hierarchy of the SM, and reduce to a 4321 model at the TeV scale.

\subsection{Extra dimension models}

\begin{figure}
\centerline{\includegraphics[width=0.6\linewidth]{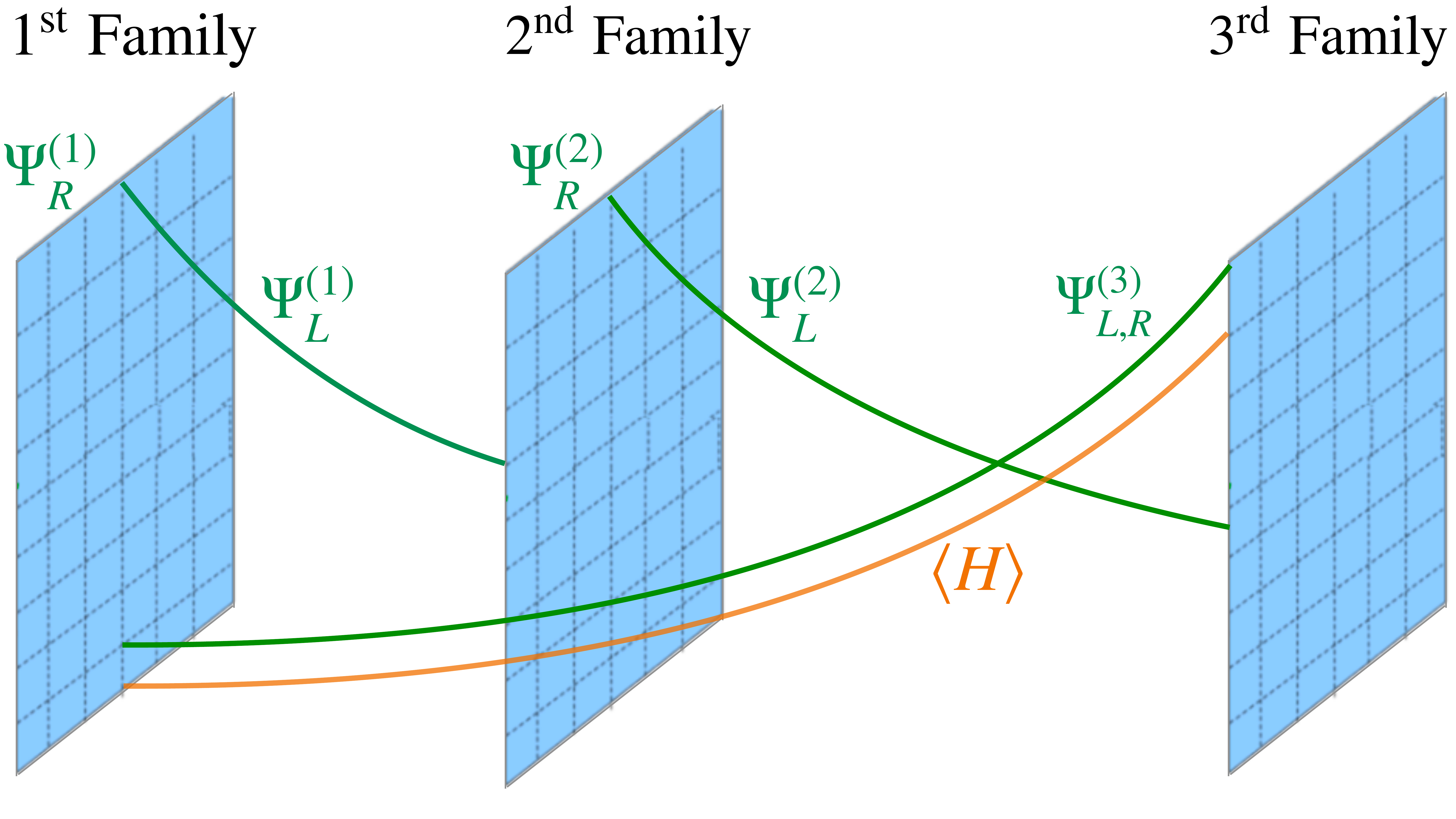}}
\caption[]{Multi-brane setup of the extra dimension models discussed in the main text, sketching the SM fermion zero-mode and Higgs profiles. The horizontal direction corresponds to the extra dimension $y$ in Eq.~\ref{eq:RSmetric}.}
\label{fig:Branes}
\end{figure}

Taking one step further, the idea of deconstructing flavor can be completed to a 5D model.\cite{Fuentes-Martin:2020pww,Fuentes-Martin:2022xnb} Fermion families would be localized differently along the extra dimension, explaining why they feel differently the breaking of PS or EW groups. A combined explanation of the flavor hierarchy and $B$-anomalies in a $U(2)$-like approach is better implemented with a Randall-Sundrum (RS) geometry,
\begin{equation}
{\rm d} s^2=-{\rm d} y^2+ e^{-2ky}\, \eta_{\mu\nu }\,{\rm d} x^{\mu}{\rm d} x^{\nu}, \label{eq:RSmetric}
\end{equation}
so the Higgs hierarchy problem can be solved à la RS. In this setup, it is necessary the introduction of extra branes to localize the SM families as shown in Fig.~\ref{fig:Branes}. The flavor structure differs from the usual flavor anarchy paradigm in RS models where the Higgs is localized in the infrared (IR) brane (the third family brane in Fig.~\ref{fig:Branes}) and Yukawa couplings are written there. The flavor hierarchy is then explained due to the different profiles of the fermions in the extra dimension. In the multi-brane setup, Yukawa couplings for each family are written on the different branes, so the hierarchy is achieved due to suppression in the 5D Higgs profile. This has the advantage that allows to localize RH fields on the branes, which suppresses RH rotations, as they can be a dangerous additional breaking of the flavor $U(2)$ symmetry.

There are several gauge groups we can consider in the 5D bulk. With a PS group,\cite{Fuentes-Martin:2020pww} the LQ appears as a KK resonance localized close to the IR brane, so it interacts strongly with the third family. This is a holographic realization of a LQ as composite resonance of a strongly coupled sector. Another possibility for the bulk gauge group is $SU(4)_h\times SU(4)_l\times G_{EW}$, where light families are charged under $SU(4)_l$ and the third family under $SU(4)_h$.\cite{Fuentes-Martin:2022xnb} This choice reintroduces a gauge LQ receiving mass from the breaking of the symmetry through boundary conditions, and allows to raise the KK scale to avoid limits from some EW observables as $Z\to \bar \tau \tau$.

\section{Conclusions}

$B$-anomalies hint New Physics coupled dominantly to the third family. A $U_1$ vector leptoquark can explain both neutral- and charge-current anomalies in a natural way, and 4321 models are its simplest UV completion that can account for the $B$-anomalies. These models suggest interesting connections with deep questions, such as the unification of quark and leptons in a non-universal way, the connection with the flavor puzzle, and the possible multi-scale origin of the hierarchies, perhaps realized via localization in an extra dimension.

\section*{Acknowledgments}

JML would like to thank L. Allwicher, G. Isidori, B. A. Stefanek and F. Wilsch for their valuable help in the preparation of this talk. The work of JML is supported by the European Research Council (ERC) under the European Union’s Horizon 2020 research and innovation program under grant agreement 833280 (FLAY).


\section*{References}

\begin{thebibliography}{99}


\bibitem{Cornella:2021sby}
C.~Cornella, D.~A.~Faroughy, J.~Fuentes-Martin, G.~Isidori and M.~Neubert,
JHEP \textbf{08}, 050 (2021)
[arXiv:2103.16558 [hep-ph]].

\bibitem{LHCb:2021trn}
R.~Aaij \textit{et al.} [LHCb],
[arXiv:2103.11769 [hep-ex]].

\bibitem{Capdevila:2020rrl}
B.~Capdevila, A.~Crivellin, C.~A.~Manzari and M.~Montull,
Phys. Rev. D \textbf{103}, no.1, 015032 (2021)
[arXiv:2005.13542 [hep-ph]].

\bibitem{Angelescu:2021lln}
A.~Angelescu, D.~Be\v{c}irevi\'c, D.~A.~Faroughy, F.~Jaffredo and O.~Sumensari,
Phys. Rev. D \textbf{104}, no.5, 055017 (2021)
[arXiv:2103.12504 [hep-ph]].

\bibitem{DiLuzio:2017vat}
L.~Di Luzio, A.~Greljo and M.~Nardecchia,
Phys. Rev. D \textbf{96}, no.11, 115011 (2017)
[arXiv:1708.08450 [hep-ph]].

\bibitem{Bordone:2017bld}
M.~Bordone, C.~Cornella, J.~Fuentes-Martin and G.~Isidori,
Phys. Lett. B \textbf{779}, 317-323 (2018)
[arXiv:1712.01368 [hep-ph]].

\bibitem{Greljo:2018tuh}
A.~Greljo and B.~A.~Stefanek,
Phys. Lett. B \textbf{782}, 131-138 (2018)
[arXiv:1802.04274 [hep-ph]].


\bibitem{Fuentes-Martin:2020bnh}
J.~Fuentes-Mart\'\i{}n and P.~Stangl,
Phys. Lett. B \textbf{811}, 135953 (2020)
[arXiv:2004.11376 [hep-ph]].

\bibitem{Fuentes-Martin:2020pww}
J.~Fuentes-Martin, G.~Isidori, J.~Pag\`es and B.~A.~Stefanek,
Phys. Lett. B \textbf{820}, 136484 (2021)
[arXiv:2012.10492 [hep-ph]].


\bibitem{Fuentes-Martin:2022xnb}
J.~Fuentes-Martin, G.~Isidori, J.~M.~Lizana, N.~Selimovic and B.~A.~Stefanek,
[arXiv:2203.01952 [hep-ph]].


\end{thebibliography}
\end{document}